# Harmonizing General Relativity with Quantum Mechanics

Antonio Alfonso-Faus[*]

*E.U.I.T. Aeronáutica, Plaza Cardenal Cisneros s/n, 28040 Madrid, Spain

**Abstract.** Gravitation is the common underlying texture between General Relativity and Quantum Mechanics. We take gravitation as the link that can make possible the marriage between these two sciences. We use here the duality of Nature for gravitation: A continuous warped space, wave-like, and a discrete quantum gas, particle-like, both coexistent and producing an equilibrium state in the Universe. The result is a static, non expanding, spherical, unlimited and finite Universe, with no cosmological constant and no dark energy. Mach´s Principle is reproduced here by the convergence of the two cosmological equations of Einstein. From this a Mass Boom concept is born given by $M = t$, M the mass of the Universe and t its age. Also a decreasing speed of light is the consequence of the Mass Boom, $c = 1/t$, which explains the Supernovae Type Ia observations without the need of expansion (nor, of course, accelerated expansion). Our Mass Boom model completely wipes out the problems and paradoxes built in the Big Bang model, like the horizon, monopole, entropy, flatness, fine tuning, etc. It also eliminates the need for inflation. Finally the relation proposed by Weinberg in 1972 is here explained in terms of a gravitational cross section for all gravitational masses.

**Keywords:** gravitation, relativity, gravity quanta, cosmology, supernovae, big bang, Mach's Principle, mass-boom.

**PACS:** 03.65. –W, 04.60. –M, 04.62. +V, 98.80. -K

## INTRODUCTION

Einstein's effort to extend the validity of classical mechanics to velocities close to the speed of light was very successful. He created Special Relativity and doing this he liberated the limits imposed on Newtonian Mechanics: it was not applicable to the case of bodies moving at a speed close to the speed of light.

Later on he continued his effort to include Gravitation in his ideas on relativity and created General Relativity, a very successful effort too. By then mechanics was on its foot, valid for small as well as high velocities, as compared with the speed of light. It was also valid for normal sizes, the see and touch world around us, as well as very large sizes, as compared with the size of the see able Universe.

But a dilemma was still puzzling the scientific community. Nature appeared to answer questions about its essence in accordance with the type of experiment set up for these questions: Wave nature? Particle nature? Both answers were affirmative if the questions were put in the proper way. Then a dual nature was clear for everybody: waves and particles were at the base of all Nature, both at the same time.

The theory that connected these two aspects of nature was Quantum Mechanics, a wave theory and a particle theory (because Heisenberg uncertainty principle is built in this wave theory) that extended the validity of mechanics to the small world of atoms and fundamental particles, the Quantum world.

Again, a big puzzle was still plaguing science: General Relativity, a theory of continuous variables, had produced the two cosmological equations to deal with the Universe as a whole. And Quantum Mechanics, the wave and particle theory of Nature, appeared to explain the world satisfactorily.

But, what about Gravitation? Only a theory of the wave-type, General Relativity, dared to treat it. No quantum treatment exists today that has a well developed theory to deal with the problem of quantum gravity.

One way to introduce a new approach to this problem is to look at the cosmological equations and try to include the wave and particle nature into them. General Relativity is clear; it is a geometrical view of gravitation where curvature plays a central role. A curved space mechanically implies a centripetal force, which is just the force of gravitation that gives an inward "push". And a gas of gravity quanta implies an outward force, the gas pressure that in fact balances the inward push. Hence we have here the picture of gravitation with a dual nature: Wave-like inward force because of the curvature of space, and particle-like, outward force due to the gas pressure of the gravity quanta.

## EINSTEIN COSMOLOGICAL EQUATIONS

The Mass-Boom concept presented elsewhere [1] implies a decreasing speed of light c, inversely proportional to cosmological time t. Then the product ct is a constant, and this can be taken as the size of the Universe, constant, with no expansion [2]. The cosmological scale factor R being constant drastically reduces the number of terms in the cosmological equations (R'=R''=O). Also the Hubble "constant" H is zero in this case, the same as the Lambda cosmological constant that equals (or is proportional to) H. Then we are left with only two terms in each of the Einstein cosmological equations, the pressure (density) term and the curvature term:

$$8\pi G \frac{p}{c^2} + \frac{Kc^2}{R^2} = 0$$
$$-\frac{8\pi}{3} G\rho + \frac{Kc^2}{R^2} = 0$$
(1)

Both equations are reduced to just one only if, and only if, the gravity quanta gas pressure p is made equal to $-1/3\, \rho\, c^2$, where $\rho$ is the density of the Universe:

$$p = -\frac{1}{3}\rho c^2$$
$$\frac{8\pi}{3} G\rho = \frac{kc^2}{R^2}$$
(2)

It is clear that this implies a model of the Universe that is spherical, k=1, closed, finite and unlimited.

The density of the Universe is equal to the mass M divided by the volume v = $2\pi^2 R^3$. Then one has

$$\frac{8\pi}{3} G \frac{M}{2\pi^2 R^3} = \frac{c^2}{R^2}$$

And using R = ct as the size of the Universe we arrive at

$$\frac{4}{3\pi} \frac{GM}{c^2} = R = ct \tag{3}$$

We can interpret this equation as a form of Mach's principle if we enunciate it as follows: The relativistic energy $mc^2$ of any mass m is of the order of its gravitational potential energy with respect to the mass M of the rest of the Universe:

$$\frac{4}{3\pi} \frac{GMm}{R} = mc^2 \tag{4}$$

The conclusion is that the two cosmological equations, which are at the base of the General Relativity approach to any model of the Universe, are just Mach's Principle. This Principle is built in the Einstein's field equations,

## THE MASS-BOOM

Presented elsewhere [1] is the idea that the Action Principle may be used to derive Einstein's field equations if the factors in front of the integrals are constant:

$$G/c^3 = \text{constant},$$
$$mc = \text{constant} \tag{5}$$

The second equation is just a manifestation of the constancy of momentum. Then from Mach Principle in the form in (3) we get:

$$M = t \tag{6}$$

in a certain system of units. This is the Mass-Boom: the equivalence of mass and time. Hence one has

$$Mc = ct = R \tag{7}$$

and one can take the size of the Universe R = ct as unity. The relation

$$c = 1/t \qquad (8)$$

explains [2] the Supernovae type Ia findings without the need of postulating any accelerating expansion of the Universe, nor the existence of dark energy (in fact the cosmological constant is also zero [3], and we are left with the initial static model of Einstein that we now know is stable). In our model there is no horizon problem nor any of the numerous problems built in the Big Bang theory which motivated the construction of inflation theories.

## A NEW COSMOLOGIAL MODEL WITH TIME-VARYING "CONSTANTS"

The Hubble red shift was interpreted as an expansion of the Universe. Here we have a Mass-Boom non-expanding model [4]and therefore we can interpret the red shift as an effect of a shrinking Quantum world [5], atoms and fundamental particles, immersed in a static, constant size Universe. This can be achieved by showing that Planck's "constant" ℏ is proportional to the speed of light c, subject to the condition of constancy of nuclear and electrical forces at the fundamental particle level.

If we impose the constancy of nuclear forces we get:

$$\frac{mc^2}{r_p} = \frac{m^2 c^3}{\hbar} = const.\frac{c}{\hbar} = const. \qquad (9)$$

Also by imposing the constancy of electrical forces we have:

$$\frac{e^2}{r_p^2} = \frac{e^2 m^2 c^2}{\hbar^2} = const.\frac{\hbar c}{\hbar^2} = const.\frac{c}{\hbar} = const. \qquad (10)$$

Then the constancy of quantum forces imposes that ℏ must be equal to c = 1/t (in a certain system of units). Then from the condition of constancy of momentum, as in (5), we have that the quantum sizes, which are of the order of the Compton wave length ℏ/mc, decrease like ℏ. But this is not the end of the story. In fact we have found arguments to believe that Planck's constant ℏ is an absolute real constant, contrary to our previous thought [5]. One argument comes from the concept of a gravitational cross section. We have shown in [6] and [7] that there is a basic concept, not very well known, that we have called the gravitational cross section σ_g of a mass m, that is of the order of the product of its gravitational radius, $2Gm/c^2$, and the gravitational radius of the Universe $2GM/c^2$, where M is the mass of the visible Universe. Its size ct is of the same order as its gravitational radius. The basic relation is then

$$\sigma_g \approx 4\frac{Gm}{c^2}\frac{GM}{c^2} \qquad (11)$$

The physical meaning of this relation is of extreme astrophysical importance. We have presented in [9] objects like the whole universe, clusters of galaxies, galaxies, globular clusters, nebulae, protostars and even a fundamental particle like the proton, all following this law, where $\sigma_g$ is of the order of the geometrical size of the object in question. If we take the Compton wave length for the proton of mass $m_p$, $\hbar/m_p c$, as the approximate geometrical size of the proton, i.e. as $\sigma_g$, then we have from (11)

$$\left(\frac{\hbar}{m_p c}\right)^2 \approx 4\frac{Gm_p}{c^2}\frac{GM}{c^2} \qquad (12)$$

In this relation, using (5), the right hand side being constant, we get the result that $\hbar$ must be a real absolute constant. Also we get a very interesting prove of Weinberg's relation [8]. Weinberg constructed a mass m from the quantities G, $\hbar$, c and $H_0$ that resulted to be not too different from the mass of a typical elementary particle. We could take it as a meaningless numerical coincidence, but the resultant combination is very much closer to a typical elementary particle mass than any other combination. It relates a single cosmological parameter, $H_0$ (the Hubble constant), to the fundamental constants G, $\hbar$, c and m. Gravitation is present here through the gravitational constant G. Quantum Mechanics through the constant $\hbar$ of Planck. And relativity through the speed of light c. We believe that this relation is of the utmost importance and provides the link between these approaches to Nature: Relativity, Gravitation, Quantum Mechanics and Cosmology. We provide here an explanation for this relation based upon a quantum gravity approach that has deep astrophysical implications.

Weinberg provided the relation between G, $\hbar$, c, $H_0$ and m as an unexplained one, as follows [8]:

$$\left(\frac{\hbar^2 H_0}{Gc}\right)^{1/3} \approx m \qquad (13)$$

If we take $H_0$ very close to $1/t$, where t is the age of the Universe, then (13) is equivalent to

$$\left(\frac{\hbar^2}{Gct}\right)^{1/3} \approx m \qquad (14)$$

Now, taking the Machean relation (3) as $2GM/c^2 \cong ct$ in (12) we get for the proton mass

$$m_p \approx \left(\frac{\hbar^2}{Gct}\right)^{1/3} \qquad (15)$$

which is precisely Weinberg's relation. Our gravitational cross section concept coincides, within a Machean approach, with Weinberg's idea. We conclude that this is an important finding in support of the quantum gravity ideas presented in [6]. And from (5) and (8) we see that the mass of the proton is proportional to t, as the mass-Boom predicts. The proportionality of all gravitational masses to the cosmological time is an effect that appears to be a general one.

## A PREDICTION

The fine structure constant α is given by

$$\alpha = \frac{e^2}{\hbar c} \qquad (16)$$

No significant time variations have been observed for this constant, in the cosmological scale of size and time. One would tend to think that if ħ and the charge e are taken as constants then the speed of light c must be a constant. However we have shown in [10] that the speed of light c does not enter into the fine structure constant. I.e., the fine structure constant, and the corrected Maxwell equations with this result, is given by

$$\alpha = \frac{e^2}{\hbar} \qquad (17)$$

Then we have that since ħ is a real absolute constant so is the charge of the electron, in electrostatic units, as above. The Zeeman displacement d is given by

$$d = \frac{e}{mc} \text{ (e in electromagnetic units)} \qquad (18)$$

In electrostatic units d is just e/m. Since e is an absolute constant here, and m follows the mass-boom law (6), d varies as c, i.e. as 1/t. Hence, by measuring time variations in this displacement we may validate this theory.

## ACKNOWLEDGEMENTS

I am grateful to all sponsors of the Eighth Symposium on Frontiers of Fundamental Physics, Madrid 16-19 October 2006, for their support, in particular to the Madrid and Valencia Technical Universities. I am also grateful to Prof. Fullana for his help in preparing the paper for the A.I.P. proceedings.